\documentclass[12pt]{article}

\usepackage{amsmath}
\usepackage{amsfonts}
\usepackage{graphicx}

\usepackage[final=true]{hyperref}

\usepackage{authblk}

\title{New explanation of Raman peak redshift in nanoparticles}

\author[1]{A.\,P. Meilakhs}
\author[1,2]{S.\,V. Koniakhin\thanks{kon@mail.ioffe.ru}}
\affil[1]{Ioffe Physical Technical Institute, St. Petersburg, 194021 Russia}
\affil[2]{St. Petersburg Academic University of the Russian Academy of Sciences, St. Petersburg, 194021, Russia}

\begin{document}
  \maketitle
\section{Abstract}
In this letter, we propose a new model that explains the Raman peak downshift observed in nanoparticles with respect to bulk materials. The proposed model takes into account discreteness of the vibrational spectra of nanoparticles. For crystals with a cubic lattice (Diamond, Silicon, Germanium) we give a relation between the displacement of Raman peak position and the size of nanoparticles. The proposed model does not include any uncertain parameters, unlike the conventionally used phonon confinement model (PCM), and can be employed for unambiguous nanoparticles size estimation.

\section{Main}
In nanoparticles the crystalline Raman peak position is downshifted and asymmetrically broadened with respect to bulk materials. Previously the phonon confinement model (PCM) was used to give qualitative and quantitative explanation of this effect. In the PCM the Gaussian-like shape envelope of the atomic displacement wave is introduced and the phonon wave function has a form
 \begin{equation}
\psi (\vec r) = A_q \exp(-\alpha r^2/ L^2) \exp(-i \vec q \vec r),
 \end{equation}
where $\vec q$ is the phonon wave vector, $L$ stands for the diameter of nanoparticles, and $\alpha$ is a dimensionless parameter that regulates strength of confinement of phonons to the center of nanoparticles. According to relation (1) the harmonics with wave vector $-q$ from the envelope allow the momentum conservation in the process of photon-phonon scattering for the phonons with wave vector $+q$. It leads to the possibility of interaction between light and phonons with $q \neq 0$ that have frequencies lower than the frequency of the phonon from the center of Brillouin zone ($q = 0$). Finally, scattering of light on the phonons with lower frequencies explains the downshift of the Raman peak position and its asymmetric broadening.

The dependence of Raman peak position on nanoparticle diameter gives possibility to employ Raman spectroscopy for nanoparticles size measurements. Experimental Raman studies followed by analysis of obtained spectra in the framework of PCM were provided previously for diamond (see \cite{NDbook} and references there), germanium \cite{Ge1, Ge2}, silicon \cite{Sibook} nanoparticles.

The PCM provided quite good agreement between predicted and measured Raman peak position and shape. Nevertheless, the following problems makes determining nanoparticles size based on PCM inconvenient. First, the suggested form of the phonon wave function (see eq. (1)) is inherently ambiguous and the value of parameter $\alpha$ is arbitrary. In an early paper on PCM \cite{Conf2} it was shown that $ \alpha = 8\pi^2$ gives the best agreement between experimental and theoretical Raman peak positions. Since then this value became default for calculating the shape of Raman peak in nanoparticles. However Irmer et al. have noticed \cite{ENN} that setting of $\alpha$ to be equal to $8\pi^2$ means that nearly 100\% of lattice vibration energy is concentrated in 2\% of nanoparticle volume, which is unrealistic. In ref. \cite{Zi} Zi \textit{et al.} via analyzing eigenmodes of sphere vibrations another value of PCM parameter is given $\alpha = 9.67$. Eq. (1) shows that particle size $L$ is a part of the combination $\frac{\alpha}{L^2}$ in the PCM and consequently its value inherently dependens on the value of $\alpha$ parameter.

The noticed uncertainty of the value of parameter $\alpha$ makes application of PCM for independent nanoparticle size determination inherently impossible.

In a nanoparticle much smaller than the phonon mean free path, the phonons undergo multiple events of reflection and scattering on the nanoparticle surface during the lifetime. Consequently, the vibrational eigenmodes of nanoparticles are not propagating plane waves but a superposition of plane waves resulting in standing waves. Precise solution, which gives the nanoparticle vibration modes and their coupling with light, is comprehensive and has been numerically solved only for small nanocrystallites. However, it is possible to qualitatively describe the discretization of nanoparticle vibrations and how the discretization affects the Raman spectra of nanoparticles.

To take the discretization into account we consider a simple model -- a one dimensional chain of $N$ atoms. Masses of atoms are denoted as $m$, interatomic distances as $a$, interaction between neighboring atoms is characterized by elastic constant $k$, both ends of the chain are free. As it can be easily verified, eigenfrequencies of this system are

 \begin{align}
\omega_s = 2\sqrt{k/m} \sin(q_s a/2),\nonumber \\
q_s = \frac{\pi s}{a N}, s=1, 2,.. N-1.
 \end{align}
Maximal frequency of vibrations tends to $2\sqrt{k/m}$ as the lengths of the chain tends to infinity, but for any finite length of the chain this frequency is never reached. Minimal and maximal possible wave vectors differ from zero and the reciprocal lattice vector $\pi/a$, correspondingly by

 \begin{equation}
\Delta q = \frac{\pi}{L+a},
 \end{equation}
where $L = (N-1) a$ is the length of the chain. There is a similar shift in chains with several types of atoms \cite{Kot}. Thus, consideration of discreteness of wave vectors leads to the downshift of maximal frequency of vibrations.

Such wave vector shift also occurs in real three dimensional structures. Due to small size of nanoparticles, this shift is not negligible. Hence maximal frequencies of vibrations in nanoparticles are less then those in bulk material. This effect completely explains the red shift of the Raman peak position in nanoparticles (Pic. 1).

\begin{figure}
\centering
\includegraphics[width=0.6\textwidth]{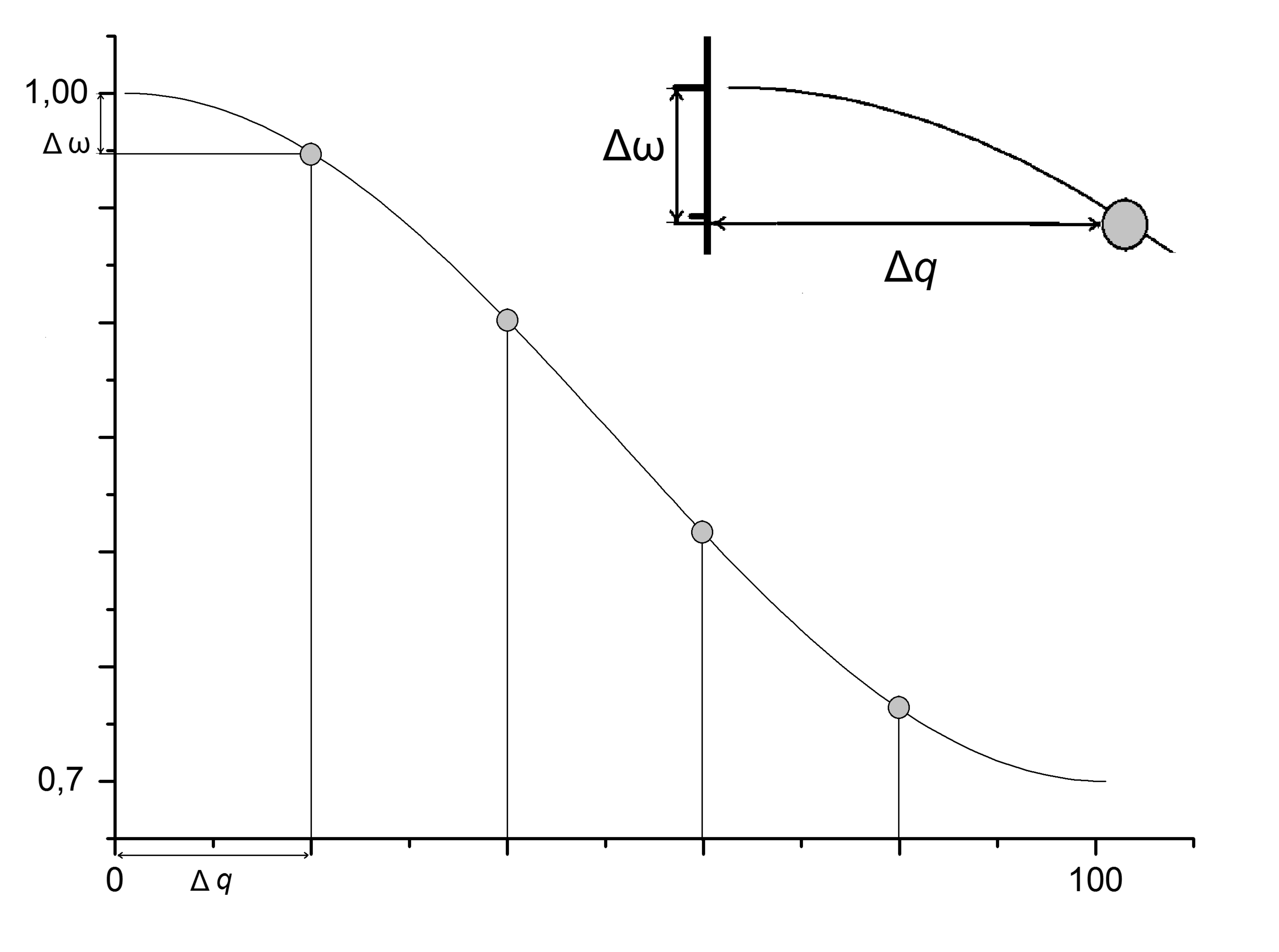}
\caption{Schematic vibrational spectrum of nanoparticle. Solid line stands for dispersion relation for the optical phonons in bulk material. Actual frequencies  of vibrations of nanoparticle are shown by circles. It was shown that the wave vector shift from the center of Brillouin zone, $\Delta q$, leads to the shift of maximal frequency of vibrations $\Delta \omega$. This is the reason why the red shift of the Raman peak occurs in nanoparticles.}
\end{figure}

We derive a formula relating the frequency shift and the diameter of the nanoparticle, for crystals with cubic lattice (diamond, germanium, silicon). We assume that the particle has approximately the shape of a rectangular parallelepiped, and the relation between the wave vector shift in each direction and the linear size of the particle in this direction is the same as in one dimensional chain:
 \begin{equation}
\Delta q_\alpha = \frac{\pi}{L_\alpha+a},
 \end{equation}
where $\alpha = x, y, z$,  $L_\alpha$ -- is a linear size of the particle in the direction $\alpha$, $a$ -- the lattice constant. We choose directions $x, y, z$ to be three forth-order axes of cubic crystal.

Linear size of a nanoparticle is an order of magnitude higher than the lattice constant hence wave vector shift is an order of magnitude smaller than the reciprocal lattice vector. So we expand the frequency shift in the Tailor series in wave vector shifts and take into account only the minor terms:
 \begin{equation}
\Delta \omega = \frac{1}{2}\frac{\partial^2 \omega}{\partial q_\alpha \partial q_\beta} \Delta q_\alpha \Delta q_\beta.
 \end{equation}
Thus, frequency shift and wave vector shifts are related by second-order symmetric tensor.  For  crystals with cubic lattice, in $\Gamma$ point of a Brillouin zone, all the principal values of such a tensor are equal, and the three fourth-order axes of crystal are the principal axes of the tensor \cite{Ans}:
 \begin{equation}
\frac{\partial^2 \omega}{\partial q_\alpha \partial q_\beta} =C \delta_{\alpha \beta}.
 \end{equation}
We substitute the expressions (4,6) into Eq.~(5) and obtain the following expression for the frequency shift:
 \begin{equation}
\Delta \omega = \frac{1}{2} C \pi^2 \left( \frac{1}{(L_x+a)^2}+\frac{1}{(L_y+a)^2}+\frac{1}{(L_y+a)^2} \right).
 \end{equation}
The dispersion law of optical phonons can be approximated by
 \begin{equation}
\omega (q) = A + B cos(qa),
 \end{equation}
where $a$ is the lattice constant. We substitute expression (8) into (6) and obtain
 \begin{equation}
C = - B a^2.
 \end{equation}
We suppose that all the linear sizes are approximately equal to the diameter $L$. Then we have
 \begin{equation}
\frac{1}{(L_x+a)^2}+\frac{1}{(L_y+a)^2}+\frac{1}{(L_z+a)^2} = \frac{3}{(L+a)^2}.
 \end{equation}
We substitute the expressions (9, 10) into equation (7) and finally we have the sought-for expression for the frequency shift
 \begin{equation}
\Delta \omega = -\frac{3 \pi^2 a^2 B }{2 (L+a)^2}.
 \end{equation}
We can also determine the particle diameter if the Raman peak shift is known
 \begin{equation}
L = \pi a \sqrt{\frac{3  B }{2 |\Delta \omega|}} - a.
 \end{equation}

There are plenty of other ways to determine the diameter of nanoparticles. For instance, the size of detonation nanodiamonds was determined by small angle X-ray scattering (SAXS) and transmission electron microscopy (TEM) \cite{DND2},  dynamic light scattering (DLS) \cite{DND3} and was acquired by computer simulations \cite{DND4}. We compare sizes determined by other methods with those determined by Raman scattering data analyzed with present formula.

In a paper by Ager at. al. \cite{Ager}, the value of constant $B$ for nanodiamonds was suggested to be $B = 91.25$ cm$^{-1}$, lattice constant of diamond $a = 0.357$ nm. Frequency shift measured in \cite{NDShend} is $4$ cm$^{-1}$ for nanodiamond with diameter $6$ nm measured by small angle neutron scattering. We substitute $\Delta \omega = 4$ cm$^{-1}$ into Eq.(12) and find the diameter of nanodiamonds to be $6.2$ nm. In \cite{NDPevts} frequency redshift for $4.5$ nm nanodiamonds is found to be 10 cm$^{-1}$. Using Eq.(11) we find that the downshift for the nanodiamond of this linear size should be $7.3$ cm$^{-1}$.

In a paper \cite{Ge2} frequency shifts are measured for germanium nanoparticles with diameters $2.6$, $3$, $9.6$ and $13$ nm, and Raman peaks are indicated at $295.9$, $296.7$, $299.4$, $299.5$ cm$^{-1}$, respectively. By approximating phonon dispersion presented in \cite{Di}, we find $B = 18.6$ cm$^{-1}$ for germanium. Raman peak for bulk germanium is at $B = 300.7$ cm$^{-1}$ \cite{Bulk}. Similarly, we find $B=31.6$cm$^{-1}$  for silicon \cite{Di}. Frequency shifts for silicon particles of different sizes are presented in Ref. \cite{Si}. Comparison of the theoretical curves, calculated with formulae (11,12) and the experimental data for the considered materials are presented in  Fig. 2.	

\begin{figure}
\centering
\includegraphics[width=1.1\textwidth]{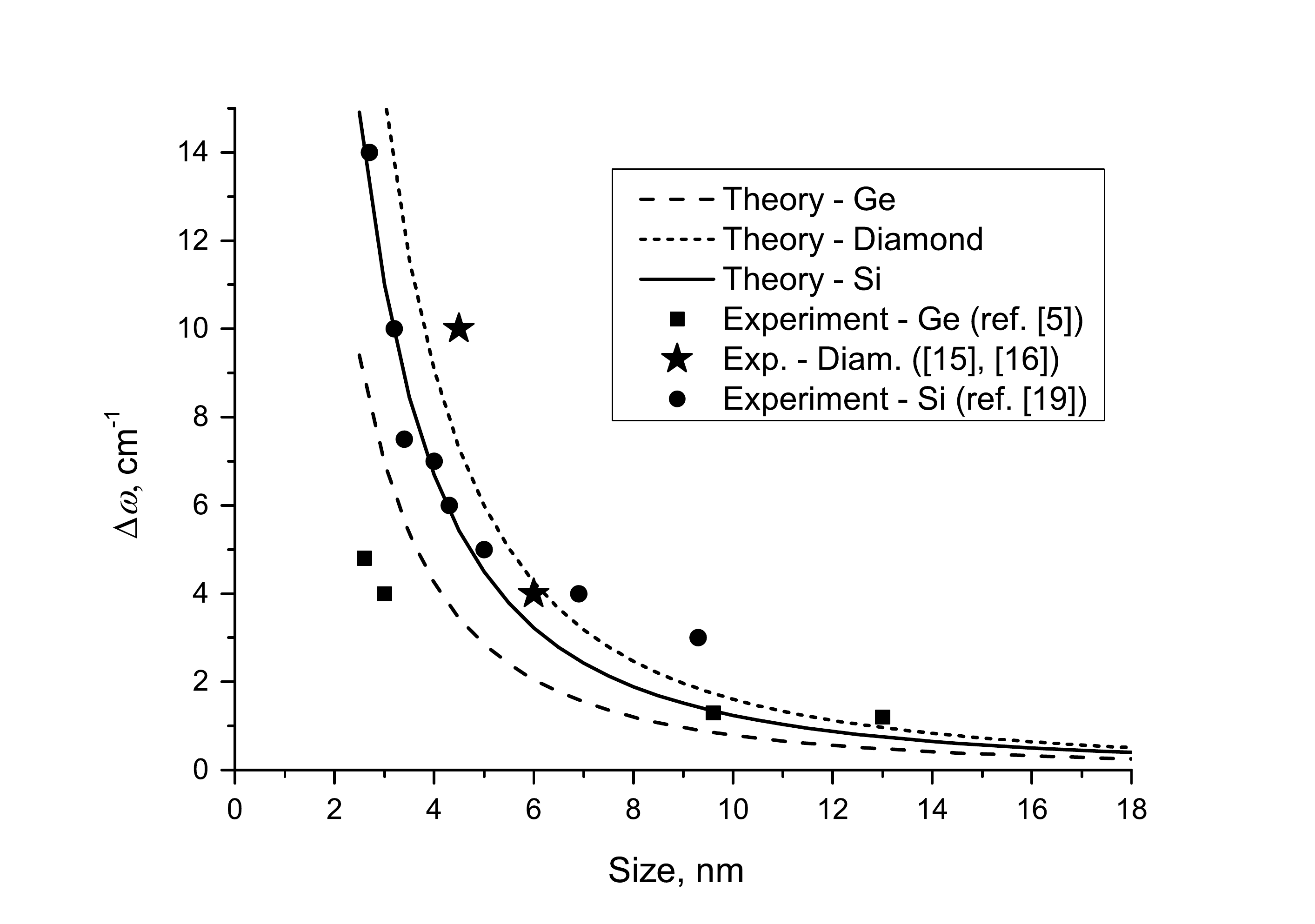}
\caption{Nanoparticle diameter vs Raman peak shifts. The long-dashed, short-dashed and solid lines are theoretical curves calculated using the Eq.(11), for germanium, diamond and silicon respectively. The experimental data from papers \cite{Ge2, NDShend, NDPevts, Si} is also shown.}
\end{figure}

The typical crystallite diameter for detonation nanodiamonds is about 3-6 nm, which is the energetically favorable size in the process of their synthesis. Therefore, it is impossible to make a comparison with experimental data for a wide range of sizes of nanodiamonds. However, it is interesting to compare the calculations by formula (12) with the calculations based on the PCM. The calculation was performed with $B = 91.25$ cm$^{-1}$. For phonon confinement model we used two values of parameter $\alpha$ which are common in literature. Results for parameter $\alpha = 8\pi^2$ are the same as in \cite{DND0} (see Fig.2 in this paper). For $\alpha = 9.67$ Raman peak shifts are sufficiently smaller, which is consistent with the calculations for silicon in \cite{Zi}. Comparison of the theoretical curves, based on the proposed model and on phonon confinement model are presented in Fig. 3. It is clear that the curve drawn by the formula (11) has the dependence $\Delta \omega \sim 1/L^2 $, for almost every crystallite sizes and deviates from it only for nanoparticles of small size, comparable to the lattice constant. The results of calculations by the formula (11) are superior to the results of the phonon confinement model with $ \alpha = 9.67 $ for about 2 times and about 3 times less then the data for $ \alpha = 8 \pi^2 $. Finally, setting $\alpha=22$ in the PCM allows to obtain Raman peak position downshift predicted by present theory.

\begin{figure}
\centering
\includegraphics[width=1.1\textwidth]{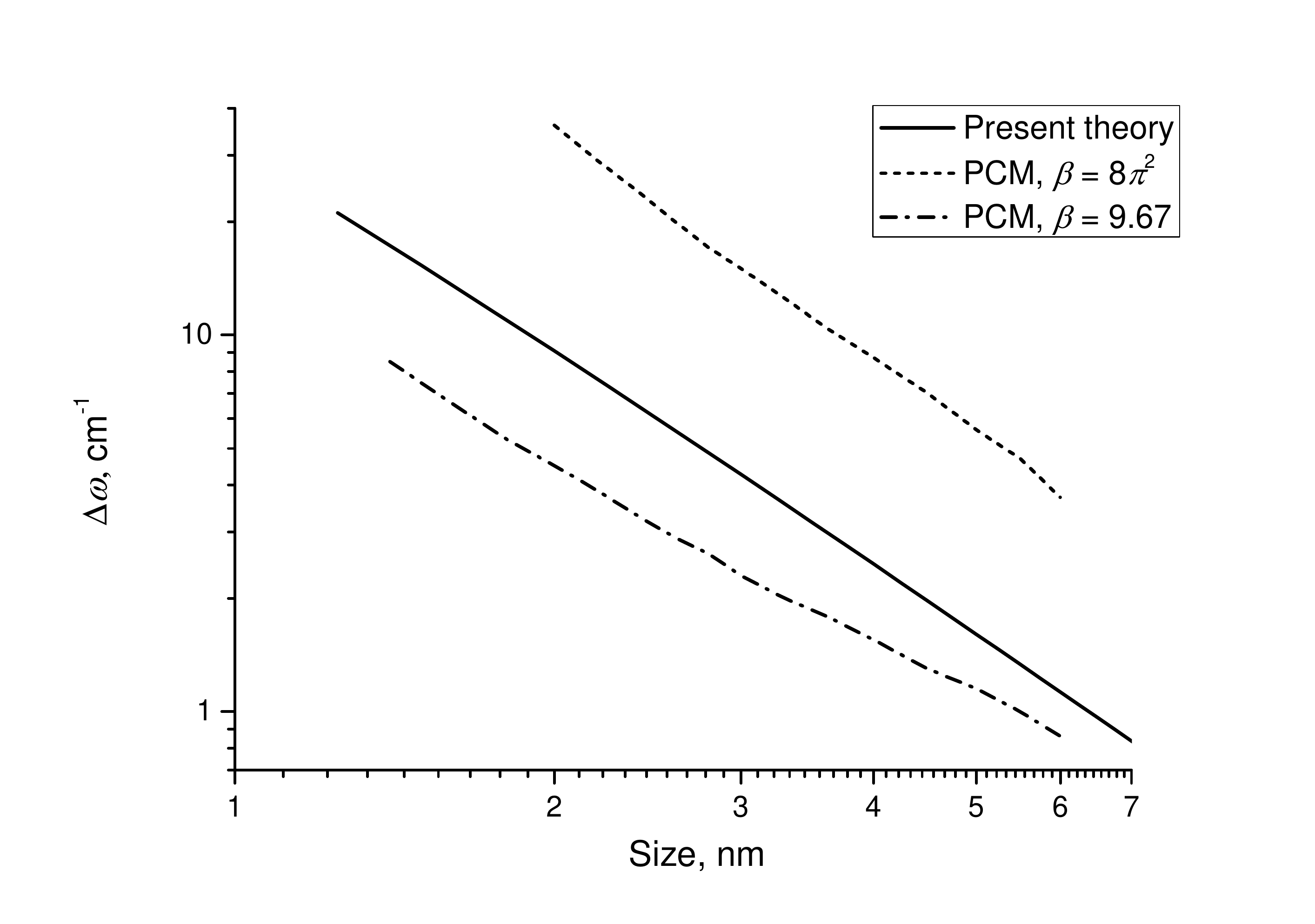}
\caption{Nanoparticle diameter vs Raman peak shifts for nanodiamonds. Parameter that characterize the phonon dispersion in diamond $B=91.25$. Results of calculations by formula (12) are drawn with solid line. Dashed lines represent the results of PCM with various values of $\alpha$ parameter. The numerical results of present theory correspond to $\alpha = 22$ parameter in the PCM.}
\end{figure}

The proposed model also explains the absence of the frequency shift for nanoparticles immersed into a matrix of other material, whereas peak broadening is still observed \cite{InMat}. Indeed, the set of wave vectors, and hence the frequency shift, strongly depends on the boundary conditions. At the same time, the difference between the nearest wave vectors that affects the broadening on boundary conditions is weak \cite{Kot}.

The developed theoretical model reproduces with high precision the position of Raman peak in nanoparticles made of various materials. However, the complete description of the Raman peak shape including its asymmetry strictly requires consideration of standing waves in nanoparticles and investigation of interaction of such vibrations with light, e.g. on the basis of the bond polarization model as it was suggested in \cite{Zi}. One expects that phonon-light coupling decrease with displacement of the phonon wave vector $q$ from the center of Brillouin zone.

In conclusion, we have proposed the new insight into vibrational properties of nanoparticles and principally new explanation of the experimentally observed redshift of the Raman peak position in nanoparticles was given. The derived formula (12) straightly relates the shift of the Raman peak position and size of the nanoparticle. In contrast with the yield of the commonly used PCM the derived expression does not contain any uncertain parameters. Comparison with experimental data shows that the developed theory reproduces the position of Raman peaks in different materials better than PCM.

The proposed approach will find various applications in nanoparticles size measurements.  Application of the developed theory for characterization of the carbon-based composites with interleaved sp$^2$ and sp$^3$ areas including the systems of nanodiamonds embedded into the graphite-like matrix is particularly promising.

Authors are gratefully indebted to E.\,D. Eidelman and A.\,Ya. Vul' for their attention to this study. S.\,V. Koniakhin acknowledges the Dynasty foundation and A.\,P. Meilakhs acknowledges Russian Science Foundation (grant \# 16-19-00075) for partial support.

\end{document}